
%
%
%
%
\magnification=1200
\voffset=0 true mm
\hoffset=0 true in
\hsize=6.5 true in
\vsize=8.5 true in
\normalbaselineskip=13pt
\def\doublespace{\baselineskip=20pt plus 3pt\message{double space}}
\def\singlespace{\baselineskip=13pt\message{single space}}
\let\spacing=\singlespace
\parindent=1.0 true cm



\newcount\equationumber \newcount\sectionumber 
\sectionumber=1 \equationumber=1               
\def\setsection{\global\advance\sectionumber by1 \equationumber=1} 
\def\numbe{{{\number\sectionumber}{.}\number\equationumber}
                            \global\advance\equationumber by1}
\def\numberit{\eqno{(\number\equationumber)} \global\advance\equationumber by1}
%
\def\numberal{(\number\equationumber)\global\advance\equationumber by1}
%
%
%
%


\def\ccf#1{\,\vcenter{\normalbaselines
    \ialign{\hfil$##$\hfil&&$\>\hfil ##$\hfil\crcr
      \mathstrut\crcr\noalign{\kern-\baselineskip}
      #1\crcr\mathstrut\crcr\noalign{\kern-\baselineskip}}}\,}
\def\scf#1{\,\vcenter{\baselineskip=9pt
    \ialign{\hfil$##$\hfil&&$\>\hfil ##$\hfil\crcr
      \vphantom(\crcr\noalign{\kern-\baselineskip}
      #1\crcr\mathstrut\crcr\noalign{\kern-\baselineskip}}}\,}

\def\small3j#1#2#3#4#5#6{\def\st{\scriptstyle} 
   \bigl(\scf{\st#1&\st#2&\st#3\cr
           \st#4&\st#5&\st#6\cr} \bigr)}


\def\ref#1{$^{#1)}$}    


\def\upa#1{\raise 1pt\hbox{\sevenrm #1}}
\def\dna#1{\lower 1pt\hbox{\sevenrm #1}}
\def\dnb#1{\lower 2pt\hbox{$\scriptstyle #1$}}
\def\dnc#1{\lower 3pt\hbox{$\scriptstyle #1$}}
\def\upb#1{\raise 2pt\hbox{$\scriptstyle #1$}}
\def\upc#1{\raise 3pt\hbox{$\scriptstyle #1$}}
\def\hprime{\raise 2pt\hbox{$\scriptstyle \prime$}}
\def\ccom{\,\raise2pt\hbox{,}}


\def\asymptotically#1{\;\rlap{\lower 4pt\hbox to 2.0 true cm{
    \hfil\sevenrm  #1 \hfil}}
   \hbox{$\relbar\joinrel\relbar\joinrel\relbar\joinrel
     \relbar\joinrel\relbar\joinrel\longrightarrow\;$}}
\def\Asymptotically#1{\;\rlap{\lower 4pt\hbox to 3.0 true cm{
    \hfil\sevenrm  #1 \hfil}}
   \hbox{$\relbar\joinrel\relbar\joinrel\relbar\joinrel\relbar\joinrel
     \relbar\joinrel\relbar\joinrel\relbar\joinrel\relbar\joinrel
     \relbar\joinrel\relbar\joinrel\longrightarrow$\;}}

\def\dal{\mathop{\sqcup\hskip-6.4pt\sqcap}\nolimits}

\catcode`@=11
\def\C@ncel#1#2{\ooalign{$\hfil#1\mkern2mu/\hfil$\crcr$#1#2$}}
\def\gf#1{\mathrel{\mathpalette\c@ncel#1}}      
\def\Gf#1{\mathrel{\mathpalette\C@ncel#1}}      

\def\gapx{\lower 2pt \hbox{$\buildrel>\over{\scriptstyle{\sim}}$}}
\def\lapx{\lower 2pt \hbox{$\buildrel<\over{\scriptstyle{\sim}}$}}

\def\nablaleft{\hbox{\raise 6pt\rlap{{\kern-1pt$\leftarrow$}}{$\nabla$}}}
\def\nablaright{\hbox{\raise 6pt\rlap{{\kern-1pt$\rightarrow$}}{$\nabla$}}}
\def\nablaboth{\hbox{\raise 6pt\rlap{{\kern-1pt$\leftrightarrow$}}{$\nabla$}}}

\def\boks#1#2{{\hsize=#1 true cm\parindent=0pt   
  {\vbox{\hrule height1pt \hbox
    {\vrule width1pt \kern3pt\raise 3pt\vbox{\kern3pt{#2}}\kern3pt
    \vrule width1pt}\hrule height1pt}}}}

\def\heading{ }
\def\range{ }

\def\body{\vfill\eject\parindent=1.0 true cm
 \ifx\spacing\singlespace\singlespace\else\doublespace\fi}
\def\title#1{\centerline{{\bf #1}}}

\def\today{\ifcase\month\or
  January\or February\or March\or April\or May\or June\or
  July\or August\or September\or October\or November\or December\fi
  \space\number\day, \number\year}
\let\date=\today
\newcount\hour \newcount\minute
\countdef\hour=56
\countdef\minute=57
\hour=\time
  \divide\hour by 60
  \minute=\time
  \count58=\hour
  \multiply\count58 by 60
  \advance\minute by -\count58

\def\sectionskip{\penalty-500\vskip24pt plus12pt minus6pt}

\def\sec{\hbox{\lower 1pt\rlap{{\sixrm S}}{\hbox{\raise 1pt\hbox{\sixrm S}}}}}
\def\section#1\par{\goodbreak\message{#1}
    \sectionskip\nobreak\noindent{\bf #1}\vskip0.3cm \noindent}

\nopagenumbers
\headline={\ifnum\pageno=\count31\frontheadline
  \else{\ifnum\pageno=0\frontheadline
     \else{{\raise 2pt\hbox to \hsize{\paperhead}}}\fi}\fi}

\footline={\centerline{\sevenbf \folio}}
\def\frontheadline{\sevenbf \hfil}
\def\paperhead{\sevenbf \heading\ \range\hfil\folio}
\newdimen\pagewidth \newdimen\pageheight \newdimen\ruleht
\maxdepth=2.2pt
\pagewidth=\hsize \pageheight=\vsize \ruleht=.5pt

\def\onepageout#1{\shipout\vbox{ 
    \offinterlineskip 
  \makeheadline
    \vbox to \pageheight{
         #1 
 \ifnum\pageno=\count31{\vskip 21pt\line{\the\footline}}\fi
 \ifvoid\footins\else 
 \vskip\skip\footins \kern-3pt
 \hrule height\ruleht width\pagewidth \kern-\ruleht \kern3pt
 \unvbox\footins\fi
 \boxmaxdepth=\maxdepth}
 \advancepageno}}

\output{\onepageout{\pagecontents}}

\count31=-1
\topskip 0.7 true cm
\doublespace
\pageno=0
\centerline{\bf Comments on Theoretical Problems in Nonsymmetric
Gravitational}
\centerline{\bf Theory}
\centerline{\bf N. J. Cornish and J. W. Moffat}
\centerline{\bf Department of Physics}
\centerline{\bf University of Toronto}
\centerline{\bf Toronto, Ontario M5S 1A7}
\centerline{\bf Canada}
\vskip 1 true in
\centerline{\bf Revised version, June 1992}
\vskip 3.5 true in
\centerline{\bf UTPT-91-33, gr-qc/9207007 }
\par\vfil\eject
\centerline{\bf Comments on Theoretical Problems in Nonsymmetric
Gravitational}
\centerline{\bf Theory}
\centerline{\bf N. J. Cornish and J. W. Moffat}
\centerline{\bf Department of Physics}
\centerline{\bf University of Toronto}
\centerline{\bf Toronto, Ontario M5S 1A7}
\centerline{\bf Canada}

\centerline{\bf Abstract}

Damour, Deser and McCarthy have claimed
that the nonsymmetric gravitational theory (NGT) is untenable due to
curvature coupled ghost modes and bad asymptotic behavior. This claim
is false for it is based on a physically inaccurate treatment of wave
propagation on a curved background and an incorrect method for
extracting asymptotic behavior. We show that the flux of gravitational
radiation in NGT is finite in magnitude and positive in sign.
\vskip 0.5 true in
The nonsymmetric gravitational theory (NGT) has been extensively studied over
a period of years$^{1,2}$ and these studies have shown that the theory
is a mathematically consistent alternative to Einstein's gravitational
theory (EGT). Other possible versions of nonsymmetric gravitational theories
$^{3,4}$ have either been shown to
possess ghost poles in the linear approximation or not to contain static
spherically symmetric solutions, which have Schwarzschild-like behavior
at large distances, unless the parameter describing the Schwarzschild mass
is forced to be negative definite$^{4}$.

The NGT Lagrangian without sources is of the form$^{2}$:
$$
{\cal L}_{NGT}=\sqrt{-g}g^{\mu\nu}R_{\mu\nu}(W)
=\sqrt{-g}g^{\mu\nu}R_{\mu\nu}(\Gamma)+{2\over 3}
(\sqrt{-g}{g^{[\nu\mu]}})_{,\nu}W_{\mu},
\numberit
$$
where
$$
W^{\lambda}_{\mu\nu}=\Gamma^{\lambda}_{\mu\nu}-{2\over 3}
\delta^{\lambda}_{\mu}W_{\nu},\quad W_{\nu}={1\over 2}
(W^{\lambda}_{\nu\lambda}-W^{\lambda}_{\lambda\nu})
=W^{\lambda}_{[\nu\lambda]}.
\numberit
$$
The empty space field equations which follow from (1) are
$$
R_{(\mu\nu)}(\Gamma)=0,
\numberit
$$
$$
R_{[\mu\nu]}(\Gamma)={2\over 3}W_{[\nu,\mu]},
\numberit
$$
$$
g_{\mu\nu,\lambda}-g_{\alpha\nu}\Gamma^{\alpha}_{\mu\lambda}
-g_{\mu\alpha}\Gamma^{\alpha}_{\lambda\nu}=0,
\numberit
$$
$$
{(\sqrt{-g}g^{[\mu\nu]})}_{,\nu}=0.
\numberit
$$
These field equations must represent 12 independent equations for the
12 independent field variables $g_{\mu\nu}$ (there exist four arbitrary
coordinate transformations: $x^{\prime\,\mu}=(\partial x^{\prime\,\mu}/
\partial x^{\alpha})x^{\alpha}$, which can be used to remove 4 of the
16 $g_{\mu\nu}$'s).

Eq.(4) can be decomposed into the two sets of equations:
$$
R_{\{[\mu\nu];\sigma\}}(\Gamma)\equiv
{R_{[\mu\nu]}(\Gamma)}_{,\sigma}+{R_{[\nu\sigma]}(\Gamma)}_{,\mu}+
{R_{[\sigma\mu]}(\Gamma)}_{,\nu}=0,
\numberit
$$
and
$$
{R_{[\mu\nu]}}(\Gamma)^{;\nu}={2\over 3}{W_{[\nu,\mu]}}^{;\nu},
\numberit
$$
where $;$ denotes covariant differentiation with respect to the connection
$\Gamma^{\lambda}_{\mu\nu}$. Eqs.(6) and (7) are constrained,
in turn, by the identities
$$
{\biggl(\sqrt{-g}g^{[\mu\nu]}\biggr)}_{,\nu,\mu}=0,
\numberit
$$
$$
\epsilon^{\mu\nu\sigma\rho}{R_{\{[\mu\nu],\sigma\}}(\Gamma)}_{,\rho}=0.
\numberit
$$
The field equations are further constrained by the
four Bianchi identities,
$$
{[\sqrt{-g}g^{\alpha\nu}G_{\rho\nu}(\Gamma)
+\sqrt{-g}g^{\nu\alpha}G_{\nu\rho}(\Gamma)]}_{,\alpha}
+[\sqrt{-g}{g^{\mu\nu}}]_{,\rho}G_{\mu\nu}(\Gamma)=0,
\numberit
$$
where $G_{\mu\nu}(\Gamma)=R_{\mu\nu}(\Gamma)-1/2g_{\mu\nu}R(\Gamma)$.

Employing Eq.(5) to eliminate $\Gamma$ in favor of $g_{\mu\nu}$, Eqs.(3),
(6) and (7) represent 18 equations for $g_{\mu\nu}$. Taking into account
the six identites (9), (10) and (11), this latter set of equations provides 12
independent
field equations for the 12 independent field variables, $g_{\mu\nu}$. At no
stage have we had to refer to the vector $W_{\mu}$. $W_{\mu}$
{\it does not describe dynamical degrees of freedom},
in keeping with the fact that it corresponds
to a Lagrange multiplier. Of course, one could use Eq.(8) to solve for
$W_{\mu}$ in terms of the previously determined $g_{\mu\nu}$ but this
would serve no useful purpose.

The field equations (3), (6) and (7) yield the following static spherically
symmetric solution:
$$
ds^2=\biggl(1-{2m\over r}\biggr)\biggl(1+{\ell^4\over r^4}\biggr)dt^2
-\biggl(1-{2m\over r}\biggr)^{-1}dr^2-r^2(d\theta^2+\hbox{sin}^2\theta
d\phi^2),
\numberit
$$
and
$$
g_{[10]}(r)={\ell^2\over r^2}.
\numberit
$$
Here, $m$ and $\ell^2$ are the two constants of integration identified with the
mass and the NGT source parameter. Thus, in NGT there are now two sources
of the {\it pure} gravitational field.  We see that the source parameter
$\ell^2$
enters into the theory in a {\it nontrivial way}, since $\ell^2$ and $m$
couple nonlinearly in $g_{00}$. The parameter $\ell^2$ has been identified
at a phenomenological microscopic level with the conserved particle number
$^{2}$.

In a weak field approximation obtained from expanding $g_{\mu\nu}$ about the
Minkowski spacetime metric, $\eta_{\mu\nu}$:
$$
g_{\mu\nu}=\eta_{\mu\nu}+\epsilon h_{\mu\nu},
\numberit
$$
where $\epsilon \ll 1$, the field equations take the form
to lowest order:
$$
\dal h_{(\mu\nu)}-{h_{(\nu\sigma),\mu}}^{,\sigma}
-{h_{(\mu\sigma),\nu}}^{,\sigma}
+h_{,\mu,\nu}=0,
\numberit
$$
$$
{h_{[\mu\beta]}}^{,\beta}=0,
\numberit
$$
$$
\dal h_{[\mu\nu]}={4\over 3}W_{[\nu,\mu]},
\numberit
$$
where $\dal =\partial^{\mu}\partial_{\mu}$ and $h=\eta^{\alpha\beta}
h_{\alpha\beta}$. We see that the symmetric part
of the field equations decouples from the
skew part, and that it is identical to that of EGT. The skew
equations take the form of Kalb-Ramond-Kimura equations$^{5}$ in a
permanently fixed gauge. The spin of $h_{[\mu\nu]}$ is $J^P=0^+$ and it is not
difficult to show {\it that there are no ghost poles} due to the existence of a
restricted gauge invariance$^{6,7}$:
$$
\delta h_{[\mu\nu]}=\epsilon_{\mu,\nu}-\epsilon_{\nu,\mu},\quad
\dal \epsilon_{\mu}-{\epsilon_{\nu,\mu}}^{,\nu}=0.
\numberit
$$

Let us define the longitudinal and transverse projection operators:
$$
P^{L}_{\mu\nu}={\partial_\mu\partial_\nu\over \dal},\quad
P^{T}_{\mu\nu}=\eta_{\mu\nu}-{\partial_\mu\partial_\nu\over \dal},
\numberit
$$
then we find that
$$
h^{TT}_{[\mu\nu]}=P^{T\alpha}_\mu P^{T\beta}_\nu h_{[\alpha\beta]},
\numberit
$$
$$
h^{LL}_{[\mu\nu]}=P^{L\alpha}_\mu P^{L\beta}_\nu h_{[\alpha\beta]}=0,
\numberit
$$
$$
h^{LT}_{[\mu\nu]}=P^{L\alpha}_{[\mu} P^{T\beta}_{\nu]} h_{[\alpha\beta]}.
\numberit
$$
Also, using the gauge condition $\partial^\alpha W_\alpha=0$, we have
$$
W^{TT}_{[\mu,\nu]}=P^{T\alpha}_\mu P^{T\beta}_\nu W_{[\alpha,\beta]}=0.
\numberit
$$
Thus, the field equations to linear order are:
$$
{h^{LT}_{[\mu\beta]}}^{,\beta}=0,
\numberit
$$
$$
\dal h_{[\mu\nu]}^{TT}=0,
\numberit
$$
and
$$
\dal h^{LT}_{[\mu\nu]}={4\over 3}W_{[\nu,\mu]}.
\numberit
$$
As in the case of the exact field equations, we see that Eqs.(24) and (25)
completely determine the $h_{[\mu\nu]}$ without reference to the Lagrange
multiplier $W_{\mu}$.

Kelly$^{4}$ and Damour, Deser and McCarthy (DDM)$^{8}$ have proposed
expanding $g_{\mu\nu}$ about a pure Einstein local vacuum background metric:
$$
g_{\mu\nu}=g_{E(\mu\nu)}+\epsilon h_{\mu\nu},
\numberit
$$
where $g_{E(\mu\nu)}$ denotes the background Einstein metric. They work at
the level of the field equations, keeping only the first order
in $h_{[\mu\nu]}$ and all orders in $g_{E(\mu\nu)}$. The resulting field
equations are$^{8}$:
$$
\bar {R}_{\mu\nu}(g_E)=0,
\numberit
$$
$$
\bar {D}^{\alpha}F_{\mu\nu\alpha}-4{{{\bar R^{\alpha}}_{\,\,\mu}}
^{\beta}}_{\nu}(g_E)
h_{[\alpha\beta]}={4\over 3}W_{[\nu,\mu]},
\numberit
$$
$$
\bar {D}^{\nu}h_{[\mu\nu]}=0,
\numberit
$$
where $F_{\mu\nu\alpha}$ and $\bar {D}^{\alpha}$ denote the cyclic curl of
$h_{[\mu\nu]}$ and the background covariant
derivative, respectively. All operations are in the background metric $g_E$
space.

As before, the equations (28)-(30) can be solved for $h_{[\mu\nu]}$ without
specifying the Lagrange multiplier $W_{\mu}$. However, in the gauge
$\partial^{\alpha}W_{\alpha}=0$, DDM proceed to take the divergence of (29)
which gives the wave equation:
$$
\bar D^{\mu}\bar D_{\mu}W_{\nu}=-3\bar D^{\mu}
({{{\bar R^{\alpha}}_{\,\,\mu}}^{\beta}}_{\nu}(g_E)h_{[\alpha\beta]}).
\numberit
$$
They argue that this is an inhomogeneous wave equation for $W_{\mu}$,
so $W_{\mu}$ has $1/r$ fall off in the wave zone. They then go on to
argue that inserting this information back into Eq.(29), drives $h_{[\mu\nu]}$
to have unsatisfactory asymptotic behaviour. This approach is incorrect.
Firstly, the source term for this wave equation is not confined to the
near zone (i.e. it is not compact), so one cannot extract a $1/r$ asymptotic
form in the usual way, instead the equation must be solved globally.
For example, the static spherically symmetric solution has $h_{[10]}
=l^{2}/r^{2}$,
${{\bar{R}}^{10}}_{\;\;\; 10}(g_E)=2m/r^{3}$ which, when inserted into (29)
or (31), gives $W_{0}=3ml^{2}/2r^{4}$, in agreement
with the exact solution. Secondly, Eq.(31) corresponds to the
redundant field equation (8) for the auxiliary field $W_{\mu}$ and so
plays no part in determining the $h_{[\mu\nu]}$. The field equations that
do in fact determine $h_{[\mu\nu]}$ are obtained by expanding Eqs.(6)
and (7):
$$
\bar {D}^{\nu}h_{[\mu\nu]}=0,
\numberit
$$
$$
\bar D^\alpha \bar D_\alpha F_{\mu\nu\sigma} - h_{[\kappa\{\mu]}
\left({{{\bar R^{\kappa}}_{\nu\sigma\}}}}^{\lambda}(g_E)\right)_{;\lambda}
+{\bar R_{\{\mu\nu}}^{\alpha\beta}(g_E){h_{[\alpha\beta]}}_{;\sigma\}}=0.
\numberit
$$

DDM go on to assert that the second term in (29) couples the background
curvature $\bar {R}$ to $h_{[\mu\nu]}$, causing
a violation of the restricted gauge invariance and thereby producing
ghost-like longitudinal modes. This assertion is false for
Eqs. (28)-(30), as they stand, do not sensibly describe wave propagation.
When studying gravitational waves propagating on a
curved background, careful attention must be paid to the physical
situation being modelled. A gravitational wave is a small ripple on the
geometry of a curved but slowly varying background. The words ``small, ripple
and slowly varying" convey an obvious physical picture, which must be correctly
modelled by the mathematics. In deriving Eq.(29) only the amplitude of the
perturbation has been controlled by taking $\epsilon \ll 1$. Implicit in the
linearisation is that the curvature induced by the perturbation can be
neglected in comparison to the background curvature. Nothing has been done
to enforce the geometrical optics condition that the background varies
more slowly than the disturbance. All these conditions can be made concrete
as follows. In terms of the decomposition (27), two
characteristic lengths, $L$ and $\lambda$, can be defined as the scales over
which the background and the wave vary,
$$
\partial g_{E(\mu\nu)}\sim {g_{E(\mu\nu)}\over L},\quad
\partial h_{\mu\nu}\sim {h_{\mu\nu}\over \lambda}.
\numberit
$$
The curvature induced by the wave is of order $(\epsilon^{2} / \lambda^{2})$,
while the background curvature is of order $(1/ L^{2})$. To be able to
consistently neglect self-gravitation requires
$$
{\epsilon\over \lambda} \ll {1\over L}.
\numberit
$$
Furthermore, to account for the distinction between the wave being a ripple
and the background being slowly varying, we demand that
$$
\delta={\lambda\over L} \ll 1,
\numberit
$$
so that we have the complete set of conditions:
$$
\epsilon \ll \delta \ll 1.
\numberit
$$
{\it One must always have $\delta \ll 1$ as well as $\epsilon \ll 1$
if the meaning of a gravitational wave is to make any sense}!$^{10}$.
If these conditions are not enforced, then the analysis is no longer
in the realm of geometric optics and notions such as local gauge
invariance become meaningless.

Indeed, the same is true in EGT when considering
gravitational waves propagating on a curved background. It is found that
the Lagrangian for the lowest order wave equation (without gauge conditions)
is not invariant under infinitesimal gauge transformations, and it is only when
the above conditions (37) are enforced that a conserved, gauge
invariant energy-momentum tensor with positive definite flux in the wave zone
is obtained$^{11}$.

Returning to NGT, and properly implementing the wave and background
decomposition, we find from equation (29) that to order $\delta$:
$$
\bar D^\alpha \bar D_\alpha h_{[\mu\nu]}={4\over 3}W_{[\nu,\mu]}.
\numberit
$$
This set of field equations together with (30) satisfy a restricted gauge
invariance and there are no longitudinal ghost-like modes.

The total energy of a system, in NGT, is given by
$$
E=\int t^{00} d^{3}x,
\numberit
$$
where $t^{\mu\nu}$ denotes the energy-momentum pseudo-tensor.
 From the conservation equations ${t^{\mu\nu}}_{,\nu}=0$, one finds that
for localized sources the total energy is conserved up to a flux of
energy carried to infinity by gravitational waves. Thus, the rate of
energy loss is given by
$$
{dE \over dt}=-R^{2} \oint t^{0i} \hat{n}_{i} d\Omega,
\numberit
$$
where the integration is over a sphere of radius $R$ in the wave zone,
and $\hat{n}_{i}$ is an outward pointing unit vector. A calculation yields
$$
{dE\over dt}=-{R^{2} \over 32\pi} \oint [({h^{TT\,(ij)}}_{,0})^2 +
({h^{TT\,[ij]}}_{,0})^2] d\Omega.
\numberit
$$
In the work of Krisher$^{9}$, contributions from $W_{\mu}$ were erroneously
kept in the radiation flux equation. These additional terms came from the
combination $4/3\delta_{ik}(h^{[0i]}W^{[j,k]}+h^{[ij]}W^{[k,0]})$ in
$t^{(0j)}$. Using equation (26) we see that this combination can be re-written
as $\delta_{ik}(h^{[0i]}\dal h_{LT}^{[kj]}+ h^{[ij]}\dal h_{LT}^{[0k]})$ which
falls off at least as $1/R^{3}$ by dint of equation (24). Thus, we see that
these terms should have been dropped along with all the other terms that
fall off faster than $1/R^{2}$.

Using Krisher's solution for $h^{[ij]}_{TT}$, given by
Eq.(4.39b) in his paper$^{9}$, a calculation shows that the second term in (41)
vanishes, and the gravitational flux is determined just by the familiar
Einstein quadrupole formula:
$$
\biggl({dE \over dt}\biggr)_{{\rm quad}}
=-\biggl<\biggl({\mu^2m^2\over r^4}\biggr){8\over 15}(12v^2-11
\dot r^2)\biggr>,
\numberit
$$
where $m$ and $\mu$ are the total mass and the reduced mass of the system,
respectively, ${\vec v}$ is the relative orbital velocity of gravitationally
bound objects, $\dot r = dr/dt$ for the orbital separation $r$, and the
angular brackets denote an average over an orbital period. Thus, there is no
dipole radiation in NGT$^{6}$ and the flux of energy carried to infinity is
positive definite. While DDM correctly point out the error in the signs of
the skew terms in Krisher's radiation flux equation, the observation is
inconsequential since none of the terms contribute to the flux.

The above arguements have now been supported by an
{\it exact} axi-symmetric gravitational wave solution in NGT$^{12}$.
The skew metric terms $h_{[\mu\nu]}$ were found to fall off as $1/r^2$
while the auxiliary vector field $W_{\mu}$ falls off like $1/r^3$ or
faster. Again, this clearly contradicts the assertions made by DDM about
bad asymptotic behavior in NGT. The gravitational wave flux in the wave zone
was found to be positive definite as in EGT.

In summary, the claim by DDM that NGT encounters problems with unphysical
longitudinal modes and with bad asymptotic behavior is incorrect. The
equations describing wave propagation in NGT about a Riemannian
background are gauge invariant, when the usual geometric optics conditions
are enforced. These are the same conditions that must be applied to
obtain sensible results in EGT. DDM's claim that the skew metric
components $h_{[\mu\nu]}$ fail to vanish asymptotically is based on
an erroneous method for extracting the asymptotic behavior of the
Lagrange multiplier $W_{\mu}$. When the six field equations for the
six components of $h_{[\mu\nu]}$ are solved (which naturally do not
refer to the Lagrange multiplier), it is found that NGT has
healthy asymptotic behavior. A direct calculation of the gravitational
energy flux in NGT for a binary system shows the flux to be finite
in magnitude and positive in sign.

\vskip 0.3 true in
{\bf Acknowledgements}
This work was supported by the Natural Sciences and Engineering Research
Council of Canada. We thank R. A. Isaacson, T. P. Krisher, M. Clayton,
P. Savaria, P. F. Kelly and R. B. Mann for helpful discussions.
\vskip 0.3 true in
\centerline{\bf References}

\item{1.}{J. W. Moffat, Phys. Rev. D{\bf 19}, 3554 (1979).}
\item{2.}{For a recent review of NGT, see: J. W. Moffat, {\it Gravitation-
A Banff Summer Institute}, eds. R. B. Mann and P. Wesson, World Scientific,
Singapore, p. 523, 1991.}
\item{3.}{P. F. Kelly and R. B. Mann, Class. Quantum Grav. {\bf 3}, 705
(1986); {\bf 4}, 1593 (1987).}
\item{4.}{P. F. Kelly, Class. Quantum Grav. {\bf 8}, 1217 (1991);
Erratum, in press.}
\item{5.}{M. Kalb and P. Ramond, Phys. Rev. D{\bf 9}, 2274 (1974); V. I.
Ogievetskii and I. V. Polubarinov, Sov. J. Nucl. Phys. {\bf 4}, 156 (1967);
T. Kimura, Prog. Theor. Phys. {\bf 64}, 351 (1980); for a corresponding
treatment of Maxwell's field equations, see: B. Lautrup, K. Danske
Vidensk. Selsk. Mat. Fys. Medd. {\bf 35}, 1 (1967);
N. Nakanishi, Prog. Theor. Phys. {\bf 35}, 1111 (1966); {\bf 38}, 881
(1967).}
\item{6.}{R. B. Mann and J. W. Moffat, J. Phys. A{\bf 14}, 2367 (1981);
Errata A{\bf 15}, 1055 (1982).}
\item{7.}{R. B. Mann and J. W. Moffat, Phys. Rev. D{\bf 31}, 2488 (1985);
T. P. Krisher and C. M. Will, Phys. Rev. D{\bf 31}, 2480 (1985).}
\item{8.}{T. Damour, S. Deser, and J. McCarthy, Phys. Rev. D{\bf 45}, R3289
(1992).}
\item{9.}{T. P. Krisher, Phys. Rev. D{\bf 32}, 329 (1985).}
\item{10.}{C. W. Misner, K. S. Thorne, and J. A. Wheeler, {\it Gravitation}
(Freeman, San Francisco, 1973) p. 956.}
\item{11.}{R. A. Isaacson, Phys. Rev. {\bf 166}, 1263 (1968); {\bf 166},
1272 (1968).}
\item{12.}{J. W. Moffat and D. C. Tatarski, ``Gravitational Waves from
an Axi-symmetric Source in the Nonsymmetric Gravitational Theory", University
of Toronto preprint, UTPT-92-01 (revised version, June 1992).}

\end